\documentclass[aps,prd,twocolumn,nofootinbib,superscriptaddress]{revtex4-1} 
\usepackage{amsmath,amssymb}
\usepackage{graphicx}
\usepackage{bigints}
\usepackage{color}

\begin{document}

\title{Tilted non-spatially-flat inflation}

\author{Bharat Ratra}
\email{ratra@phys.ksu.edu}
\affiliation{Department of Physics, Kansas State University,  Manhattan, Kansas 66506, USA}

\date{\today}

\begin{abstract}

We construct non-linear inflaton potential energy densities that describe
not-necessarily very-slowly-rolling closed and open inflation models, and 
compute tilted primordial spatial inhomogeneity power spectra that follow 
from quantum mechanical fluctuations during inflation in these models. 
Earlier non-flat inflation model power spectra computations assumed an 
inflaton potential energy density with a linear slope that resulted in 
very-slow-roll during inflation and untilted power spectra. These new 
tilted power spectra differ from those that have previously been used to 
study cosmological data in non-flat cosmological models.

\end{abstract}

\maketitle
\section{Introduction}
\label{Intro}

If general relativity provides an adequate description of gravity on 
cosmological scales --- and there is no strong evidence indicating otherwise 
--- dark energy is the dominant contributor to the current cosmological 
energy budget and powers the observed late-time accelerating cosmological 
expansion. Earlier on, prior to a redshift $z \sim 0.75$, non-relativistic 
(cold dark and baryonic) matter was the dominant contributor to the energy 
budget and was responsible for the observed earlier-time decelerating 
cosmological expansion. The simplest model consistent with these observations 
is the flat $\Lambda$CDM model \citep{Peebles1984}, the current ``standard'' 
cosmological model. In this model spatial hypersurfaces are chosen to be flat
and the dark energy is the cosmological constant $\Lambda$, with the next 
biggest contributor to the current cosmological energy budget being cold 
dark matter (CDM). For reviews see Refs.\ \citep{CosmoRev}.

The flat $\Lambda$CDM model is consistent with a variety of observational 
constraints, including cosmic microwave background (CMB) anisotropy 
observations \citep{Planck2020}, baryon acoustic oscillation (BAO) data 
\citep{eBOSS2020}, Hubble parameter ($H(z)$) measurements 
\citep{Farooqetal2016}, and Type Ia supernova (SNIa) apparent magnitude 
observations \citep{Scolnicetal2018}. The standard model is also consistent 
with more recent constraints from probes of the intermediate redshift 
Universe, that include data between $z \sim 2.3$ of the highest redshift 
BAO observations and $z \sim 1100$ of the CMB data. However, the intermediate 
redshift data constraints are not yet as restrictive as the lower-redshift 
BAO, $H(z)$, and SNIa ones, nor as restrictive as the higher-redshift CMB
anisotropy ones. These intermediate-redshift constraints include those from
HII galaxy 
apparent magnitude versus redshift data \citep{HII}, angular size as a 
function of redshift measurements  \citep{AngSize}, quasar X-ray and UV 
flux observations \citep{QSO}, and gamma-ray burst data \citep{GRB}. 
      
While most current measurements are not inconsistent with the spatially-flat
$\Lambda$CDM standard model, they also do not rule out mildly curved spatial
hypersurfaces or, weakly varying in time and space, dynamical dark energy.
Near-future measurements are anticipated to provide significantly more 
restrictive constraints that should help distinguish between the options 
and better determine cosmological parameter values \citep{nearfuture}.

There are however some suggestions of inconsistencies between observations
and the standard flat $\Lambda$CDM model. For example, differences between 
(model-dependent) measurements of the Hubble constant $H_0$ could be an 
indication of a problem with the standard model.\footnote{For recent
reviews see Ref.\ \citep{CosmoRev}.} 
An early median statistics estimate, \citep{ChenRatra2011}, 
$H_0 = 68 \pm 2.8$ km s$^{-1}$ Mpc$^{-1}$, is consistent with a number of 
more recent $H_0$ measurements made using a variety of methods \citep{H0}, 
including from CMB anisotropy data, 
$H_0 = 67.36 \pm 0.54$ km s$^{-1}$ Mpc$^{-1}$ \citep{Planck2020}.
On the other hand, some local measurements of the expansion rate favor 
a value significantly larger than the CMB one, 
$H_0 = 73.2 \pm 1.3$ km s$^{-1}$ Mpc$^{-1}$ 
\citep{Riessetal2021}.\footnote{We note that some other local expansion
rate determinations of $H_0$ are slightly lower and have larger error
bars \citep{localH0}.} Similar issues affect measurements of other 
parameters but the difference between $H_0$ measurements is the most
significant. For reviews of these issues see Refs.\ \citep{CosmoRev}.

Given such potential inconsistencies, and given the improving quality and 
amount of data, there is significant interest in studying cosmologies that 
have a free parameter or two more than the flat $\Lambda$CDM model. A 
widely considered option is 
dynamical dark energy that mildly varies in time and space. Scalar field 
dynamical dark energy ($\phi$CDM) models are a popular example \citep{phiCDM}.
Allowing for non-zero spatial curvature is another option now under study. 
For recent discussions of observational constraints on spatial curvature and 
dark energy dynamics from a variety of different data sets, see Refs.\ 
\citep{curvatureconstraints, DEconstraints}.\footnote{Compared to the 
cosmological constant, dynamical dark energy density evolves more similarly 
to spatial curvature energy density and this results in weaker constraints 
on both new parameters when compared to the case when either only non-zero 
spatial curvature or only dark energy density dynamics is assumed 
\citep{timevaryingnonflatconstraints}.} For recent discussions of non-flat
cosmological models, see Refs.\ \citep{Baumgartner}. 

CMB anisotropy data provide the most restrictive constraints on cosmological 
models. To use these data to constrain cosmological parameters 
of a model requires knowing the primordial power spectrum of spatial 
inhomogeneities as a function of wavenumber for the model. In the inflation
scenario \citep{InflationGuth, InflationOthers} quantum mechanical zero-point 
fluctuations in the inflaton field during inflation generate the spatial 
inhomogeneities \cite{InflationInhomogeneities, Fischleretal1985}. 
If inflation lasts for a long time spatial curvature is redshifted away to
insignificance
(and this is by far the most commonly considered case). In this case, if the 
inflaton slow-rolls down a relatively-flatter inflaton potential energy 
density the cosmological scale factor grows exponentially in time (this is 
spatially-flat de Sitter inflation) and the resulting primordial power 
spectrum is close to scale invariant, \citep{HPYZ}, with very little 
tilt. It is possible to increase the power spectral tilt by choosing an 
inflaton potential energy density that causes the inflaton to evolve more 
rapidly during inflation and makes the scale factor grow only as a 
power of time (this is spatially-flat power-law inflation) 
\citep{LucchinMatarrese1985, Ratra1992, Ratra1989}. 

Gott \citep{Gott1982} generalized inflation to the open cosmological model.
In this open-bubble inflation model a spatially-open bubble nucleates and 
the interior inflates for only a limited time so as to not redshift away all 
spatial curvature. If necessary, an earlier, pre open-bubble nucleation, 
epoch of less-limited spatially-flat inflation can be used to produce 
spatial homogeneity. Alternately, a slow enough open-bubble nucleation 
process might ensure that the interior of the open bubble is sufficiently 
spatially homogeneous.

Hawking's prescription for the initial quantum state of the universe 
\citep{Hawking1984} --- that the functional integral include only those 
field configurations which are regular on the Euclidean section --- 
suggests that the universe nucleated as a closed de Sitter Lanczos 
(inflation) model on the Lorentzian section \citep{Hawking1984, Ratra1985}.
The equator of the Euclidean (de Sitter Lanczos) four sphere is identified
with the waist of the Lorentzian de Sitter Lanczos hyperboloid, which is
where the nucleation occurs 
\citep{Hawking1984, Ratra1985}.\footnote{For variants of this 
picture see Refs.\ \citep{Linde, Grattonetal2002, LasenbyDoran2005, Gordonetal2021}.} 
A slow enough nucleation process might ensure a sufficiently spatially 
homogeneous closed inflating de Sitter Lanczos model. Again, inflation 
can occur for only a limited amount of time so as to not redshift away all
spatial curvature.\footnote{For a discussions of the nucleation of
open and closed cosmological models, see Refs.\ \citep{Cespedesetal2020} and
references therein.}

To compute the power spectrum of spatial inhomogeneities generated by quantum 
fluctuations during inflation in a given model requires the solution of the 
spatial inhomogeneities linear perturbation equations and a set of initial 
conditions. Spatially flat, open, and closed de Sitter spacetimes have as 
large a symmetry group as Minkowski spacetime; spatially-flat power-law 
inflation and the non-flat inflation models we study here have less symmetry
than the flat, open, and closed de Sitter inflation models. The initial 
condition prescription we utilize here give inflaton scalar field two-point 
functions in flat and closed de Sitter inflation models that have the 
symmetries of those spacetimes, \citep{Ratra1985, Fischleretal1985, Ratra2017}. 
This initial condition prescription has also been 
used to compute the primordial power spectrum in the spatially-flat power-law
inflation model \citep{Ratra1989, Ratra1992}, giving a tilted primordial
power spectrum that is in agreement with the result of Ref.\ 
\citep{LucchinMatarrese1985}. All these inflation model spacetimes, as well
as the two we consider in this paper, are conformally flat, i.e., when 
expressed in terms of conformal time their line-elements are proportional to
the Minkowski spacetime line-element.

The initial condition prescription we use here is that during inflation, 
for large wavenumbers, inside the horizon, at early time, the conformally
rescaled inflaton scalar field modes, as a function of conformal time, should
be quantum-mechanically normalized simple harmonic oscillators.

In the open-bubble inflation model this initial condition prescription
\citep{Ratra1994} results in a late-time energy density inhomogeneity 
power spectrum \citep{RatraPeebles1995, RatraPeebles1994} that is the 
generalization to the open inflation case 
\citep{LythStewart1990}\footnote{Lyth and Stewart \citep{LythStewart1990}
also use this initial condition prescription in their more approximate  
computation of the wavenumber dependence of the power spectrum} of the 
scale-invariant spectrum of the flat model \citep{HPYZ}. In a variant 
of the open-bubble inflation model, that includes an initial epoch of 
spatially-flat de Sitter inflation, applying the initial condition 
prescription in the first epoch and following the computation through 
the bubble nucleation process, results in a primordial power spectrum 
that is observationally indistinguishable from that of the case when 
the initial conditions are applied inside the second, open-bubble epoch, 
\citep{BucherYamamotoetal1995, Gorskietal1998}.\footnote{For other early 
papers on open inflation, see Refs.\ \citep{OpenInflation}.}  

This initial condition prescription has also been used in a computation of 
the primordial power spectrum in the closed de Sitter inflation model 
\citep{Ratra1985, Ratra2017}. In this case this initial condition 
prescription is equivalent to Hawking's \citep{Hawking1984} prescription 
of only including field configurations regular on the Euclidean section
\citep{Ratra1985}. It also leads to a de Sitter invariant ground state 
inflaton scalar field two-point correlation function \citep{Ratra1985}.
The wavenumber dependence of the resulting primordial power spectrum  
is the generalization of the scale-invariant spectrum in the spatially-flat 
case \citep{HPYZ} to the closed universe 
\citep{closedscaleinvariant}.\footnote{There are other computations of 
primordial spectra in the closed de Sitter inflation model \citep{LasenbyDoran2005, Massoetal2008, AsgariAbbassi2015, Bongaetal2016, Handley2019, KieferVardanyan2021}, using different initial 
conditions compared to our prescription. We emphasize that the initial 
conditions we use in the closed de Sitter case results in a scalar field 
perturbation two-point function that is de Sitter invariant \citep{Ratra1985}.} 

The open and closed inflation model primordial power spectra computations
of Refs.\ \citep{RatraPeebles1995, Ratra2017} were done in models that 
had an inflaton ($\Phi$) potential energy density $\propto (1 - \epsilon\Phi)$,
where $\epsilon$ is a small constant. These are very-slow-roll models and so 
the resulting power spectra are untilted on small scales for infinitesimal 
$\epsilon$. When these power spectra were used in a non-flat $\Lambda$CDM 
model analysis of the Planck 2015 CMB anisotropy data \citep{Planck2016} 
it was found that these data favored closed spatial geometry 
\citep{Oobaetal2018, ParkRatra2019}, even in combination with BAO, $H(z)$, 
SNIa, and other non-CMB data, where these data jointly favored 
about a 1\% spatial curvature energy density contribution to the cosmological 
energy budget at 5$\sigma$ significance \citep{ParkRatra2020}. 

A more correct analysis of the CMB anisotropy data requires tilted open and 
closed model primordial power spectra. Such spectra would be generated by 
quantum mechanical inflaton fluctuations in open and closed inflation models 
with non-linear inflaton potential energy densities, unlike the linear 
potential energy density function assumed in the analyses of Refs.\ 
\citep{RatraPeebles1995, Ratra2017}. Pending such a computation, data 
analyses have been performed utilizing the primordial power 
spectra of Refs.\ \citep{RatraPeebles1995, Ratra2017} multiplied by $k^{n - 1}$,
where $k$ is the wavenumber and $n$ is the power spectral index (with $n = 1$ 
being the scale-invariant case in the flat model) \citep{LesgourguesTram2014}.
Using this phenomenological primordial power spectrum to define a tilted 
non-flat $\Lambda$CDM model for the analysis of CMB anisotropy data, 
in this model the Planck 2018 data \citep{Planck2020} favors positive 
spatial curvature contributing about 1\% to the cosmological energy budget 
at 1.6$\sigma$, but when BAO data are added to the mix the result is 
consistent with flat spatial hypersurfaces \citep{Planck2020}; a similar 
result was originally found from the Planck 2015 data \citep{Planck2016}. 
It is of interest to determine whether the power spectra 
\citep{LesgourguesTram2014} used in these analyses 
\citep{Planck2016, Planck2020} can be generated by inflaton quantum 
fluctuations in non-flat non-very-slow-roll inflation models that are 
closed but very close to flat, deviating from flatness at only the 
$\sim 1$\% level. A recent numerical study in closed inflation models that
computes power spectra generated for a few different initial conditions finds
that it is possible to generate spectra of the form assumed in Refs.\
\citep{Planck2016, Planck2020}, \citep{Guthetal2022}, at least in the closed
case.
          
Here we consider open and closed inflation models with non-linear inflaton
potential energy densities. In this paper we generalize the exponential
potential energy density of the flat-space power-law inflation model 
\citep{LucchinMatarrese1985, Ratra1992, Ratra1989} to inflaton potential energy
densities that allow for not-necessarily very-slow-roll inflation in open and
closed models. In the 
very-slow-roll limit these potential energy densities reduce to that 
$\propto (1 - \epsilon\Phi)$ used in the open and closed inflation models of
Refs.\ \citep{RatraPeebles1995, Ratra2017}, while at large $\Phi$ they become 
the exponential potential energy density used in the spatially-flat power-law 
inflation model of Refs.\ \citep{LucchinMatarrese1985, Ratra1992, Ratra1989}.
Here we are interested in a non-very-slow-roll limit of these models, and
potentially in the parameter-space range where they deviate from spatial 
flatness at the $\sim 1$\% level.

Our computed primordial power spectra of spatial inhomogeneities --- that 
result from quantum mechanical zero-point fluctuations during the inflation 
epoch in these tilted closed and open models --- differ from power spectra 
that have previously been used to analyze observational data in closed and 
open cosmological models. These power spectra have been used in the analyses
of CMB anisotropy and other data, \citep{deCruzetal2022}. It is interesting
that there appears to be some additional ambiguity in the form of non-flat
inflation model power spectra, caused by the ambiguity in the form of the
assumed non-flat inflation initial conditions, compared to what happens in
the flat inflation case.
   
In Section II we summarize the background geometry of the closed and open 
models and the Einstein-scalar-field model equations of motion. For more 
detailed descriptions see Refs.\ \citep{RatraPeebles1995, Ratra2017}. In 
Section III we determine the inflaton potential energy densities we use 
and solve the spatially homogeneous background equations of motion in the 
inflation epoch of the closed and open models. We solve the linear 
perturbation equations in Section IV, where we compute the late-time 
primordial power spectra during inflation in the closed and open models.
We conclude in Section V. Appendix
A list some results in the spatially-flat tilted inflation model, that we
use for comparison to some of our results on smaller scales during inflation 
in the non-flat models when spatial curvature is unimportant. Appendices
B and C describe primordial power spectra definitions and conventions in the 
flat, closed, and open models.

\section{Technical Preliminaries}
\label{technicalpreliminaries}

\subsection{Spatially homogeneous background geometries}
\label{spatiallyhomogeneous}

The positive spatial curvature (closed) FLRW model has line element
\begin{eqnarray}
\label{closedlineelement}
    ds^2 & = & dt^2 - a^2(t) H_{ij} (\vec x) dx^i \, dx^j \\ 
         & = & dt^2 - a^2(t) \left [ d\chi^2
          + {\rm sin}^2\!(\chi) \left\{d\theta^2 + {\rm sin}^2\!
          (\theta ) \, d\phi^2 \right\} \right], \nonumber
\end{eqnarray}
where $a(t)$ is the cosmological scale factor, $H_{ij}(\vec x)$ is the metric on
the closed spatial hypersurfaces, the `radial' coordinate $0 \le \chi < 
\pi$, and $\theta , \phi$ are the usual angular coordinates on the two-sphere. 
The square of the distance between two points, $(t, \chi , \theta , \phi )$
and $(t, \chi' , \theta' , \phi' )$, is
\begin{eqnarray}
   \sigma^2  =  2 a^2 (t) 
              \left[ - 1 + {\rm cos}(\gamma_3) \right],
\end{eqnarray}
\begin{eqnarray}
\label{closedgamma3}
   {\rm cos}(\gamma_3)  =  {\rm cos}(\chi) {\rm cos}(\chi')
     + {\rm sin}(\chi) {\rm sin}(\chi') {\rm cos}(\gamma_2),
     \cr
\end{eqnarray}
where $\gamma_2$ is the usual angle between the two points $(\theta , \phi)$
and $(\theta' , \phi')$ on the two-sphere,
\begin{eqnarray}
\label{gamma2}
   {\rm cos}(\gamma_2) = {\rm cos}(\theta) {\rm cos}(\theta')
     + {\rm sin}(\theta) {\rm sin}(\theta') {\rm cos}(\phi - \phi').
\end{eqnarray}

The negative spatial curvature (open) FLRW model has line element
\begin{eqnarray}
\label{openlineelement}
    ds^2 & = & dt^2 - a^2(t) H_{ij} (\vec x) dx^i \, dx^j \\ 
         & = & dt^2 - a^2(t) \left [ d\chi^2
          + {\rm sinh}^2\!(\chi) \left\{d\theta^2 + {\rm sin}^2\!
          (\theta ) \, d\phi^2 \right\} \right], \nonumber
\end{eqnarray}
where $H_{ij}(\vec x)$ is now the metric on the open spatial hypersurfaces, 
and $\chi \ (0 \le \chi < \infty), \theta ,$ and $\phi$ are defined above. 
The square of the distance between two points, $(t, \chi , \theta , \phi )$
and $(t, \chi' , \theta' , \phi' )$, is
\begin{eqnarray}
   \sigma^2  =  2 a^2 (t) 
              \left[ 1 - {\rm cosh}(\gamma_3) \right],
\end{eqnarray}
\begin{eqnarray}
\label{opengamma3}
   {\rm cosh}(\gamma_3)  =  {\rm cosh}(\chi) {\rm cosh}(\chi')
     - {\rm sinh}(\chi) {\rm sinh}(\chi') {\rm cos}(\gamma_2),
     \cr
\end{eqnarray}
and $\gamma_2$ is defined in Eq.\ (\ref{gamma2}).

\subsection{Einstein-scalar-field model conventions}
\label{einsteinscalarfieldconventions}

The Einstein-scalar-field action, for metric tensor $g_{\mu\nu}$ and 
inflaton $\Phi$, is
\begin{eqnarray}
   & { } &  S  = \\
   & { } & {m_p{}^2 \over 16\pi} \int dt \, d^3\! x 
           \sqrt{-g} \left[ -R + {1 \over 2}
           g^{\mu\nu} \partial_\mu\Phi \partial_\nu\Phi -
           {1 \over 2} V(\Phi) \right] . \nonumber 
\end{eqnarray}
Here $m_p = G^{-1/2}$ is the Planck mass and $V$ is the scalar field potential 
energy density. Varying, we find the inflaton and gravitation equations of 
motion,
\begin{eqnarray}
\label{scalareom}
    & {} & {1\over \sqrt {-g}} \partial_\mu \left( \sqrt {-g} g^{\mu\nu}
      \partial_\nu \Phi \right) + {1 \over 2} V'(\Phi) = 0 ,
\end{eqnarray}
\begin{eqnarray}
\label{einsteinscalareom}
    & {} & R_{\mu\nu} = {8\pi \over m_p{}^2} \left( T_{\mu\nu}
      - {1\over 2} g_{\mu\nu} T \right) ,
\end{eqnarray}
where a prime denotes a derivative with respect to $\Phi$ and 
$T$ is the trace of the stress-energy tensor
\begin{eqnarray}
\label{scalarstresstensor}
   & { } & T_{\mu\nu} = \nonumber \\
   & { } & {m_p{}^2 \over 16\pi} \left[ \partial_\mu \Phi
           \partial_\nu\Phi - {1\over 2} g_{\mu\nu} \left\{
           g^{\lambda\rho} \partial_\lambda\Phi \partial_\rho\Phi
           - V(\Phi) \right\}\right] .
\end{eqnarray}

To derive the equations of motion for the spatially homogeneous background
fields and for the spatial inhomogeneities, we linearize Eqs.\ 
(\ref{scalareom}) 
-- (\ref{scalarstresstensor}) about an open or closed FLRW model and a 
spatially homogeneous scalar field. We work in synchronous gauge, with line 
element
\begin{eqnarray}
\label{pertlineelement}
   ds^2 = dt^2 - a^2(t) \left[ H_{ij} (\vec x) - h_{ij} (t, \vec x)
          \right] dx^i dx^j ,
\end{eqnarray}
where the background metric on the closed [open] spatial hypersurfaces, 
$H_{ij}$, is given in Eq.\ (\ref{closedlineelement}) 
[Eq.\ (\ref{openlineelement})], and the metric perturbations are 
denoted by $h_{ij}$. The expansion for the scalar field is
\begin{eqnarray}
   \Phi (t, \vec x) = \Phi_b (t) + \phi (t, \vec x) ,
\end{eqnarray}
where $\Phi_b$ and $\phi$ are the spatially homogeneous and 
inhomogeneous parts of the inflaton field (the inflaton perturbation
$\phi$ should not be confused with the angular variable $\phi$ of
Sec.\ \ref{spatiallyhomogeneous}).

\section{Tilted Closed and Open Inflation Models and Spatially Homogeneous Background Solutions}
\label{tiltedclosedopeninflationmodels}

The Einstein-scalar-field model equations of motion for the spatially 
homogeneous background fields, derived in Sec.\ III.A of Ref.\ 
\citep{RatraPeebles1995} for the open case and in Sec.\ III.A of Ref.\ 
\citep{Ratra2017} for the closed case, are 
\begin{eqnarray}
\label{scalarfieldKG}
   \ddot\Phi_b + 3 {\dot a \over a} \dot\Phi_b + {1\over 2} 
                   V'(\Phi_b) = 0 ,
\end{eqnarray}
\begin{eqnarray}
\label{scalarfieldfriedmann1}
   \left({\dot a \over a}\right)^2 = {1\over 12} \left[ \dot
                   \Phi_b{}^2 + V(\Phi_b) \right] + {\kappa^2 \over a^2} ,
\end{eqnarray}
\begin{eqnarray}
\label{scalarfieldfriedmann2}                   
   {\ddot a \over a } = - {1\over 6} \dot\Phi_b{}^2 + 
                    {1\over 12} V(\Phi_b) , 
\end{eqnarray}
where an overdot denotes a derivative with respect to time and 
$\kappa^2 = + 1 (-1)$ for open (closed) spatial hypersurfaces.

Motivated by the power-law expansion inflation model in the 
spatially-flat case, \citep{LucchinMatarrese1985, Ratra1989, Ratra1992}, for 
suitable scalar field potential energy densities, discussed below, the 
background Friedmann equation ({\ref{scalarfieldfriedmann1}) during 
closed or open inflation becomes
\begin{eqnarray}
\label{scalarfieldfriedmann1new}
   \left({\dot a \over a}\right)^2 = {Q \over a^q} + {\kappa^2 \over a^2} , 
\end{eqnarray}
where $Q$ and $q$ are constants and $0< q < 2$ for inflation. In the closed 
case where $\kappa^2 = - 1$, we require, at the waist at $t = t_i$, 
the initial condition $\dot a(t_i) = 0$, so the right hand side of 
Eq.\ ({\ref{scalarfieldfriedmann1new}}) must also vanish at the waist, which 
results in $Q a_i^p = 1$ where $a_i = a(t_i)$ and $p = 2-q$. In the closed 
case the inflation 
model includes only the $t \ge t_i$ part of the spacetime.

The integral of Eq.\ (\ref{scalarfieldfriedmann1new}) is
\begin{eqnarray}
\label{at}
 \sqrt{\kappa^2} (t - t_0) & = & a\, {}_2F_1(1/2, 1/p; 1 + 1/p; -Qa^p/\kappa^2) \\
                    & { } & - a_0\, {}_2F_1(1/2, 1/p; 1 + 1/p; -Qa_0^p/\kappa^2) , \nonumber
\end{eqnarray}
where ${}_2F_1$ is the Gauss hypergeometric function, see Ch.\ 15 of Ref.\
\citep{AS}, and $a_0 = a(t= t_0)$ is the constant of integration.

In the open case, this is
\begin{eqnarray}
\label{atopen}
  t = a\, {}_2F_1(1/2, 1/p; 1 + 1/p; -Qa^p) + {\rm constant},
\end{eqnarray}
and in the closed case, where $t_i$ is at the waist,
\begin{eqnarray}
\label{atclosed}
 t - t_i &  =  & i a\, {}_2F_1(1/2, 1/p; 1 + 1/p; Qa^p) \nonumber \\
         & { } & - i Q^{-1/p} \sqrt{\pi} {\Gamma (1 + 1/p) 
                            \over \Gamma (1/2 + 1/p)}, 
\end{eqnarray}
where $\Gamma$ is the Euler Gamma function.  

In the flat limit where $\kappa^2 \rightarrow 0$, or 
$Q a^p/\kappa^2 \rightarrow \infty$, Eq.\ (\ref{at}) becomes 
$a \propto (t - t_0)^{2/q}$, the usual flat-space tilted power-law inflation 
result, see
Eq.\ (2.5) of Ref.\ \citep{Ratra1992}. In the $q \rightarrow 0$ limit 
Eqs.\ (\ref{atopen}) and (\ref{atclosed}) reduce to the correct open and 
closed slow-roll untilted de Sitter inflation relations,  
$a(t) = {{\rm sinh} (\sqrt{Q}(t-t_0))}/\sqrt{Q}$ and 
$a(t) = {{\rm cosh} (\sqrt{Q}(t-t_0))}/\sqrt{Q}$, see Eq.\ (4.10) of Ref.\ 
\citep{RatraPeebles1995} and Eq.\ (105) of Ref.\ \citep{Ratra2017}. 

In the open case, conformal time
\begin{eqnarray}
\label{tildetopen}
  {\tilde t} = {1 \over p \sqrt{\kappa^2}} {\rm ln} 
     \left[{ \sqrt{Qa^p/\kappa^2 + 1} - 1 \over
             \sqrt{Qa^p/\kappa^2 + 1} + 1 }
     \right],
\end{eqnarray}
where $-\infty \le {\tilde t} \le 0$, while in the closed case, conformal time 
\begin{eqnarray}
\label{tildetclosed}
  {\tilde t} = {2 \over p \sqrt{-\kappa^2}} {\rm tan}^{-1} 
     \left[\sqrt{Qa^p/(-\kappa^2) - 1}\right],
\end{eqnarray} 
where $ 0 \le {\tilde t} \le \pi/p$.

With the scalar field potential energy density in the open case,
\begin{eqnarray}
\label{openped}
 V(\Phi) = {2 (6-q) Q \over (\kappa^2/ Q)^{q/p} \left[{\rm sinh} \left\{ p \Phi /{\sqrt{8q}} \right\}\right]^{2q/p}}
\end{eqnarray}
and in the closed case,
\begin{eqnarray}
\label{closedped}
 V(\Phi) = {2 (6-q) Q \over (-\kappa^2/ Q)^{q/p} \left[{\rm cosh} \left\{ p \Phi /{\sqrt{8q}} \right\}\right]^{2q/p}},
\end{eqnarray}
the homogeneous part of the scalar field equation of motion 
(\ref{scalarfieldKG}) and the Friedmann equation 
(\ref{scalarfieldfriedmann1})\footnote{More precisely, Eqs.\ (\ref{openaPhi}) and (\ref{closedaPhi}) reduce the Friedmann equation (\ref{scalarfieldfriedmann1}) to Eq.\ (\ref{scalarfieldfriedmann1new}) which is solved by Eqs.\ (\ref{atopen}) and (\ref{atclosed}).}
are satisfied by, in the open case,
\begin{eqnarray}
\label{openaPhi}
 a = (\kappa^2/ Q)^{1/p} \left[{\rm sinh} \left\{ p \Phi_b /{\sqrt{8q}} \right\}\right]^{2/p},
\end{eqnarray}
and in the closed case,
\begin{eqnarray}
\label{closedaPhi}
 a = (-\kappa^2/ Q)^{1/p} \left[{\rm cosh} \left\{ p \Phi_b /{\sqrt{8q}} \right\}\right]^{2/p}.
\end{eqnarray}
In the closed case we have used the initial condition $\Phi_b(t_i) = 0$.

In the untilted, slow-roll, small $q$ limit, both scalar field potential energy 
densities, Eqs.\ (\ref{openped}) and (\ref{closedped}), become 
$\propto (1 - \epsilon \Phi)$, where $\epsilon = \sqrt{q/2}$, 
which are the potential energy densities used in untilted very-slow-roll open 
\citep{RatraPeebles1995} and closed \citep{Ratra2017} inflation model 
computations. However. the $a(\Phi_b)$ equations (\ref{openaPhi}) 
and (\ref{closedaPhi}) appear to not behave sensibly at $q = 0$ and so it 
appears that these models do not make sense at $q = 0$. For small $q$ the 
potential energy densities, Eqs.\ (\ref{openped}) and  (\ref{closedped}), 
change only slowly with $\Phi$ and at $q = 0$ they are flat.
At $q = 0$ the scalar field will not move if is initially at rest, but $a(t)$
grows, resulting in a breakdown of Eqs.\ (\ref{openaPhi}) and 
(\ref{closedaPhi}) at $q = 0$. 

In the limit that $\Phi$ is large, both scalar field potential energy 
densities, Eqs.\ (\ref{openped}) and (\ref{closedped}), become 
$\propto {\rm exp}(-\sqrt{q/2} \Phi)$, which is the potential energy 
density used in the standard flat-space tilted inflation model
\citep{LucchinMatarrese1985, Ratra1989, Ratra1992}.

\section{Linear Scalar Perturbations During Inflation}

\subsection{Synchronous gauge linear scalar perturbation equations}

The scalar parts of the synchronous gauge inflation-epoch linear perturbation 
equations in spatial momentum space are derived in Secs.\ II and III.A of 
Refs.\ \citep{RatraPeebles1995, Ratra2017} for the open and closed cases.

With $- A ( A + 2 )$, integer $A = 0, 1, 2 \cdots$ ($A \ge 2$ modes are 
physical), being the closed model spatial Laplacian eigenvalue, and 
$- (A^2 +1)$, $A > 0$, being the open model spatial Laplacian eigenvalue, 
we define $k^2 = A^2+1$ and $\bar k^2 = A^2+4$ for the open case and 
$k^2 = A(A+2)$ and $\bar k^2= (A-1)(A+3)$ for the closed case.  

The linear scalar perturbation equations for the spatial momentum space 
scalar field $\phi (A, t)$, trace of the metric perturbation $h (A, t)$ 
(the perturbation to the size of the proper volume element), 
and the trace-free part of the metric perturbation ${\cal H} (A, t)$ 
(the shearing perturbation of the volume element) modes are
\begin{eqnarray}
\label{scalarpert1s}
     \ddot\phi + 3 {\dot a \over a} \dot\phi + {k^2 \over a^2}
                   \phi + {1\over 2} V''(\Phi_b) \phi = {1\over 2}
                   \dot h \dot\Phi_b , 
\end{eqnarray}
\begin{eqnarray}
\label{scalarpert2s}
     \ddot h + 2 {\dot a \over a} \dot h = 2 \dot\Phi_b\dot\phi -
                   {1\over 2} V'(\Phi_b) \phi ,
\end{eqnarray}
\begin{eqnarray}
\label{scalarpert3s}
     \dot{\cal H} = {k^2 \over \bar k^2} 
                   \left[ {3\over 2} \dot\Phi_b\phi - \dot h \right] ,
\end{eqnarray}
\begin{eqnarray}
\label{scalarpert4s}
     \ddot h + 6 {\dot a \over a} \dot h 
                & + & {(\kappa^2 + \bar k^2) \over a^2} h + \ddot {\cal H} 
                      + 3 {\dot a \over a} \dot {\cal H} \nonumber \\
                & + & {(\kappa^2 + \bar k^2) \over a^2} {\cal H} 
                       + {3\over 2} V'(\Phi_b) \phi = 0 ,
\end{eqnarray}
\begin{eqnarray}
\label{scalarpert5s}
     \ddot {\cal H} + 3 {\dot a \over a} \dot {\cal H} 
                 - {k^2 \over 3 a^2} {\cal H} 
                 - {k^2 \over 3 a^2} h = 0 . 
\end{eqnarray}

\subsection{Synchronous gauge linear scalar perturbation solutions}

Using Eqs.\ (\ref{scalarfieldfriedmann1new}) and 
(\ref{openped})---(\ref{closedaPhi}), Eqs.\ (\ref{scalarpert1s}) and 
(\ref{scalarpert2s}) can be re-expressed as 
\begin{eqnarray}
\label{scalarpert1a}
   & { } & [Qa^p + \kappa^2] {d^2\!\phi \over da^2}  
     + {1 \over 2a} [(8 - q)Qa^p + 6\kappa^2] {d\phi \over da} + {k^2 \over a^2} \phi + \\
   & { } & \   
      {(6 - q) \over 4 a^2} [2q Qa^p + (2 + q)\kappa^2] \phi
   = {\sqrt{2qQ} \over 2 a^{q/2}} [Qa^p + \kappa^2]^{1/2} {dh \over da}, \nonumber 
\end{eqnarray}
and
\begin{eqnarray}
\label{scalarpert2a}
 & { } & [Qa^p + \kappa^2] {d^2\!h \over da^2}  
     + {1 \over 2a} [(6 - q)Qa^p + 4\kappa^2] {dh \over da} = \\
 & { } & \ {2 \sqrt{2qQ} \over a^{q/2}} [Qa^p + \kappa^2]^{1/2} 
     {d\phi \over da} 
     + {(6 - q) \sqrt{2qQ} \over 2 a^{(2+q)/2}} [Qa^p + \kappa^2]^{1/2} \phi. \nonumber 
\end{eqnarray}
These can be combined to give a third order equation for $\phi$
\begin{eqnarray}
\label{scalarpert3rdorder}
   & { } & [Qa^p + \kappa^2] {d^3\!\phi \over da^3}  
     + {1 \over 2a} [2(8 - q)Qa^p + (10 + q) \kappa^2] {d^2\!\phi \over da^2} \nonumber \\
   & { } & \   + {1 \over 4a^2} [(48 - 10q - q^2)Qa^p + (12 - q)(2 + q) \kappa^2 + 4 k^2] {d\phi \over da} \nonumber \\
   & { } & \   + {q \over 8a^3} [2 p (6 - q) Qa^p + (6 - q)(2 + q) \kappa^2 + 4 k^2] \phi \nonumber \\
   & { } & \ = 0. 
\end{eqnarray}
Changing variables from $\phi$ and $a$ to $f$ and $x$ where $\phi = f/a^{q/2}$ 
and $x = Qa^p/\kappa^2$, this equation becomes
\begin{eqnarray}
\label{Zdiffeq}
   & { } & x^2(x+1) {d^2\!Z \over dx^2}  
     + \left[{11 \over 2} x^2 + 4 x\right] {dZ \over dx} + \\
   & { } &  
     \left[{(40 - 41 q + 10q^2)  \over 2 p^2} x + {(11 - 8 q + 2 q^2 + k^2/\kappa^2) \over p^2}\right] Z = 0, \nonumber 
\end{eqnarray}
where $Z(x) = df/dx$. The general solution $Z(x)$ of Eq.\ (\ref{Zdiffeq}) can 
be expressed in terms of Gauss hypergeometric functions and this can be 
integrated once with respect to $x$ to get $f(x)$ and so $\phi (a)$.

Defining $A_\pm$, $B$, $D$, and $G_\pm$,
\begin{eqnarray}
\label{Apmformula}
4p A_\pm = 3p -2 W \pm (2+q),
\end{eqnarray}
\begin{eqnarray}
\label{Bformula}
p B = p - W,
\end{eqnarray}
\begin{eqnarray}
\label{Dformula}
p D = p + W,
\end{eqnarray}
\begin{eqnarray}
\label{Gpmformula}
4p G_\pm = 3p + 2W \pm (2+q),
\end{eqnarray}
where
\begin{eqnarray}
W = \sqrt{-8 - 4q + q^2 - 4 k^2/\kappa^2},
\end{eqnarray}
the scalar field solution of the linear perturbation equations is
\begin{eqnarray}
\label{scalarfieldperturbationsolution}
a^{q/2} \phi & = & \tilde{c} \\
       & + & \tilde{c}_+ x^{(B-2)/2} {}_3F_2(A_+, A_-, B/2-1; B, B/2; -x) \nonumber \\
       & + & \tilde{c}_- x^{(D-2)/2} {}_3F_2(G_+, G_-, D/2-1; D, D/2; -x) . \nonumber
\end{eqnarray}
Here $\tilde{c}$ and $\tilde{c}_\pm$ are constants of integration, 
the $\tilde{c}$ solution is a gauge solution corresponding to the remnants of 
time translation invariance in synchronous gauge, and ${}_3F_2$ is a 
generalized hypergeometric function, see Ch.\ IV of Ref.\ \citep{EHTF} and 
Ch.\ 16  of Ref.\ \citep{OLBC}.

In terms of $x$, Eq.\ (\ref{scalarpert2a}) is
\begin{eqnarray}
\label{hequation}
 & { } & x^2 (x + 1) {d^2\!h \over dx^2}  
     + \left[{(8 - 3q) \over 2 p} x^2 + {(3 - q) \over p} x \right] {dh \over dx} = \\
 & { } & \ {2 \sqrt{2q} \over p} x^{3/2} (x + 1)^{1/2} {d\phi \over dx} 
     + {(6 - q) \sqrt{2q} \over 2 p^2} x^{1/2} (x + 1)^{1/2}\phi . \nonumber 
\end{eqnarray}
Using Eq.\ (\ref{scalarfieldperturbationsolution}) to evaluate the right hand 
side of this equation, it can be solved to give the trace of the metric 
perturbation solution
\begin{eqnarray}
\label{hsolution}
h  & = & c_2 - 2 c_1 (-1)^{1/p} \sqrt{x+1}\, {}_2F_1(1/2, 1+1/p; 3/2; x+1) \nonumber \\
   & + & {3 \sqrt{2q} \over p} \left({Q\over\kappa^2}\right)^{q/2p} \tilde{c} (-1)^{1/p} \sqrt{x+1}\nonumber \\
   &   &  \ \ \ \ \times {}_2F_1(1/2, 1/p; 3/2; x+1) \nonumber \\
   & + & {3 \sqrt{2q} \over 2p} \left({Q\over\kappa^2}\right)^{q/2p} \tilde{c}_+ \int dx {x^{(B-2)/2-1/p} \over \sqrt{x+1}} \nonumber \\   
   & {} & \ \ \times \big\{ {}_3F_2(A_+, A_-, B/2-1; B, B/2; -x) \nonumber \\
   & {} & \ \ \ \ \ + {(B-2) \over 3B}  {}_3F_2(A_+, A_-, B/2; B, B/2+1; -x) \big\} \nonumber \\
   & + & {3 \sqrt{2q} \over 2p} \left({Q\over\kappa^2}\right)^{q/2p} \tilde{c}_- \int dx {x^{(D-2)/2-1/p} \over \sqrt{x+1}} \nonumber \\   
   & {} & \ \ \times \big\{ {}_3F_2(G_+, G_-, D/2-1; D, D/2; -x) \\
   & {} & \ \ \ \ \ + {(D-2) \over 3D}  {}_3F_2(G_+, G_-, D/2; D, D/2+1; -x) \big\}, \nonumber
\end{eqnarray}
where the integrals can be done (and expressed as infinite series) but this is 
not of use to us and so these series are not recorded here. In 
Eq.\ (\ref{hsolution}) $\tilde{c}$ 
and $\tilde{c}_\pm$ are the constants in 
Eq.\ (\ref{scalarfieldperturbationsolution}) and $c_1$ and 
$c_2$ are constants of integration, with $c_2$ corresponding to a gauge mode.

In terms of $x$, Eq.\ (\ref{scalarpert3s}) is
\begin{eqnarray}
     {d{\cal H}\over dx}  = {k^2 \over \bar k^2} 
                   \left[ {3\sqrt{2q}\over 2 p} {\phi \over \sqrt{x(x + 1)}} - {dh \over dx} \right] .
\end{eqnarray}
Using Eqs.\ (\ref{scalarfieldperturbationsolution}) and (\ref{hsolution}) to 
evaluate the right hand side of this equation, it can be solved to give the 
trace-free metric perturbation solution
\begin{eqnarray}
\label{Hsolution}
&  & {\bar k^2 \over k^2} {\cal H} =  \\
&  & c_3 + 2 c_1 (-1)^{1/p} \sqrt{x+1}\, {}_2F_1(1/2, 1+1/p; 3/2; x+1) \nonumber \\
   & - & {\sqrt{2q} \over 2p} \left({Q\over\kappa^2}\right)^{q/2p} \tilde{c}_+ {(B-2) \over B} \int dx {x^{(B-2)/2-1/p} \over \sqrt{x+1}} \nonumber \\   
   & {} & \ \ \ \ \times {}_3F_2(A_+, A_-, B/2; B, B/2+1; -x) \nonumber \\
   & - & {\sqrt{2q} \over 2p} \left({Q\over\kappa^2}\right)^{q/2p} \tilde{c}_- {(D-2) \over D} \int dx {x^{(D-2)/2-1/p} \over \sqrt{x+1}} \nonumber \\   
   & {} & \ \ \ \ \times {}_3F_2(G_+, G_-, D/2; D, D/2+1; -x), \nonumber
\end{eqnarray}
where the integrals can be done (and expressed as infinite series) but this not
of use to us and so these series are not recorded here. In Eq.\ 
(\ref{Hsolution}) $\tilde{c}_\pm$ and $c_1$ are the 
constants in Eqs.\ (\ref{scalarfieldperturbationsolution}) and 
(\ref{hsolution}), and $c_3$ is a  constant of integration corresponding to 
a gauge mode.

Using the solutions given in Eqs.\ (\ref{scalarfieldperturbationsolution}), 
(\ref{hsolution}) and (\ref{Hsolution}), the left-hand sides of Eqs.\ 
(\ref{scalarpert4s}) and (\ref{scalarpert5s}) are proportional. Requiring they 
vanish results in the relation
\begin{eqnarray}
\label{c1tildec}
c_1 = \sqrt{2 \over q} {1 \over p} \left({Q\over\kappa^2}\right)^{q/2p} \tilde{c} {\bar k^2 \over \kappa^2}.
\end{eqnarray}
We are unable to analytically establish that the coefficients of 
$\tilde{c}_\pm$ --- the gauge-invariant contributions --- vanish in these 
equations, as they must. However, we are convinced that they indeed do vanish
since, as discussed below, the complete numerical solution (for given 
sets of parameter values) of the linear 
perturbation equations results in power spectra that are in very good 
agreement with the analytic power spectra we have derived. Additionally,
we show below that on smaller scales when spatial curvature is unimportant,
the primordial power spectra in these tilted non-flat inflation models are 
identical to the primordial power spectrum in the tilted flat model of 
Refs.\ \citep{LucchinMatarrese1985, Ratra1992, Ratra1989}, where in the
computation of Ref.\ \citep{Ratra1992} it was shown that the corresponding 
coefficients of $\tilde{c}_\pm$ vanished in the corresponding equations.

\subsection{Gauge-invariant variables solutions}

For scalar perturbations there are two independent gauge-invariant 
variables, invariant under the remnants of general coordinate 
invariance in synchronous gauge \citep{Ratra1991}. We choose these to be
\begin{eqnarray}
  &  &  \Delta_\Phi = \nonumber \\
  &  &  \ { 1\over \dot\Phi_b{}^2 + V(\Phi_b)} \left[
                   2\dot\Phi_b\dot\phi + V'(\Phi_b)\phi + 6 {\dot a\over a}
                   \dot\Phi_b\phi\right] , \\
  &  &  A_\Phi = \nonumber \\
  &  & \ {1 \over \dot\Phi_b{}^2 + V(\Phi_b) }
              \left[ 2 \dot\Phi_b\dot\phi + V'(\Phi_b)\phi 
              - \dot\Phi_b{}^2 (h + {\cal H}) \right] . 
\end{eqnarray}
Another useful gauge-invariant combination is
\begin{eqnarray}
\label{zetadef}
      {\mathcal{R}}_\Phi = {\dot\Phi_b^2 + V(\Phi_b) \over 6 \dot\Phi_b^2}
      \left[ \Delta_\Phi - A_\Phi \right] .
\end{eqnarray}

During inflation, from the $\phi$, $h$, and ${\cal H}$ solutions of the 
previous sub-section, the gauge-invariant variables are
\begin{eqnarray}
\label{DeltaPhi}
\Delta_\Phi & = & {p \sqrt{2q} \over 12} \left({Q\over\kappa^2}\right)^{q/2p} \sqrt{x+1} \\
       & \times & \Big[ (B-2) \tilde{c}_+ x^{(B-2)/2 - 1/p} {}_2F_1(A_+, A_-;B; -x) \nonumber \\
       & { }  & + (D-2) \tilde{c}_- x^{(D-2)/2 - 1/p} {}_2F_1(G_+, G_-;D; -x) \Big], \nonumber
\end{eqnarray}
and
\begin{eqnarray}
\label{APhi}
 A_\Phi & = & -{q \over 6} \left({Q\over\kappa^2}\right)^{q/2p} \sqrt{x+1} \\
       & \times & \Big[ {6 \over\sqrt{2q}} \tilde{c}_+ x^{(B-2)/2 - 1/p} \nonumber \\
       & { } & \ \ \ \ \ \ \times {}_3F_2(A_+, A_-,B/2 -1; B, B/2; -x) \nonumber \\
       & { } & + {6 \over\sqrt{2q}} \tilde{c}_- x^{(D-2)/2 - 1/p} \nonumber \\ 
       & { } & \ \ \ \ \ \ \ \times {}_3F_2(G_+, G_-,D/2 -1; D, D/2; -x) \nonumber \\
       & { } & - \left({3q \over 2} {\kappa^2 \over \bar{k}^2} x + 1 \right) {p \over\sqrt{2q}} (B-2) \tilde{c}_+ x^{(B-2)/2 - 1/p} \nonumber \\
       & { } & \ \ \ \ \ \ \times {}_2F_1(A_+, A_-;B; -x) \nonumber \\
       & { } & - \left({3q \over 2} {\kappa^2 \over \bar{k}^2} x + 1 \right) {p \over\sqrt{2q}} (D-2) \tilde{c}_- x^{(D-2)/2 - 1/p} \nonumber \\ 
       & { } & \ \ \ \ \ \ \times {}_2F_1(G_+, G_-;D; -x) \nonumber \\
       & { } & - {3 \over 2} {\kappa^2 \over \bar{k}^2} \sqrt{2q} {(B-2) \over B} \tilde{c}_+ x^{B/2 - 1/p} \nonumber \\ 
       & { } & \ \ \ \ \ \ \times {}_3F_2(A_+, A_-,B/2; B, B/2 + 1; -x) \nonumber \\
       & { } & - {3 \over 2} {\kappa^2 \over \bar{k}^2} \sqrt{2q} {(D-2) \over D} \tilde{c}_- x^{D/2 - 1/p} \nonumber \\ 
       & { } & \ \ \ \ \ \ \times {}_3F_2(G_+, G_-,D/2; D, D/2 + 1; -x)\Big]. \nonumber
\end{eqnarray}
We assume that the gauge-invariant solutions, those proportional to 
$\tilde{c}_\pm$, obey Eqs.\ (\ref{scalarpert4s}) and (\ref{scalarpert5s}), and 
use these and Eq.\ (\ref{c1tildec}) to simplify the expression for $A_\Phi$ 
to that given in Eq.\ (\ref{APhi}), see the discussion around 
Eq.\ (\ref{c1tildec}).

From the expressions for $\Delta_\phi$ and $ A_\phi$ above, 
\begin{eqnarray}
\label{zetaPhi}
{\mathcal{R}}_\Phi & = & \left({Q\over\kappa^2}\right)^{q/2p} \sqrt{x+1} \\
       & \times & \Big[ {1 \over\sqrt{2q}} \tilde{c}_+ x^{(B-2)/2 - 1/p} \nonumber \\
       & { } & \ \ \ \ \ \ \times {}_3F_2(A_+, A_-,B/2 -1; B, B/2; -x) \nonumber \\
       & { } & + {1 \over\sqrt{2q}} \tilde{c}_- x^{(D-2)/2 - 1/p} \nonumber \\ 
       & { } & \ \ \ \ \ \ \ \times {}_3F_2(G_+, G_-,D/2 -1; D, D/2; -x) \nonumber \\
       & { } & - {\sqrt{2q}p \over 8} {\kappa^2 \over \bar{k}^2} (B-2) \tilde{c}_+ x^{B/2 - 1/p} \nonumber \\
       & { } & \ \ \ \ \ \ \times {}_2F_1(A_+, A_-;B; -x) \nonumber \\
       & { } & - {\sqrt{2q}p \over 8} {\kappa^2 \over \bar{k}^2} (D-2) \tilde{c}_- x^{D/2 - 1/p} \nonumber \\ 
       & { } & \ \ \ \ \ \ \times {}_2F_1(G_+, G_-;D; -x) \nonumber \\
       & { } & - {\sqrt{2q} \over 4} {\kappa^2 \over \bar{k}^2} {(B-2) \over B} \tilde{c}_+ x^{B/2 - 1/p} \nonumber \\ 
       & { } & \ \ \ \ \ \ \times {}_3F_2(A_+, A_-,B/2; B, B/2 + 1; -x) \nonumber \\
       & { } & - {\sqrt{2q} \over 4} {\kappa^2 \over \bar{k}^2}  {(D-2) \over D} \tilde{c}_- x^{D/2 - 1/p} \nonumber \\ 
       & { } & \ \ \ \ \ \ \times {}_3F_2(G_+, G_-,D/2; D, D/2 + 1; -x)\Big]. \nonumber
\end{eqnarray}

\subsection{Initial conditions and constants of integration}

Defining $\hat{c}_\pm = \tilde{c}_\pm\sqrt{\kappa^2/Q}$, and ignoring the first, 
gauge-dependent, term on the right hand side of Eq.\ 
(\ref{scalarfieldperturbationsolution}), a more convenient form of the
inflaton field perturbation solution is
\begin{eqnarray}
\label{newscalarfieldperturbationsolution}
a \phi & = & \\
       & { } & \hat{c}_+ x^{(B-1)/2} {}_3F_2(A_+, A_-, B/2-1; B, B/2; -x) \nonumber \\
       & + & \hat{c}_- x^{(D-1)/2} {}_3F_2(G_+, G_-, D/2-1; D, D/2; -x). \nonumber
\end{eqnarray}
 
We assume as an initial condition that the conformally-rescaled scalar field 
perturbation ($a \phi$) is in the conformal time harmonic oscillator ground 
state, when non-flat inflation initiated, on small length scales. I.e., we 
require
\begin{eqnarray}
\label{initialcondition}
    {\rm lim}_{A \rightarrow \infty} a \phi(A,t) =
    \left({16\pi\over m_p{}^2}\right)^{1/2} {e^{-iA\tilde t} \over \sqrt{2A}},
\end{eqnarray}
where conformal time $\tilde t$ is defined, for the open and closed models, 
in Eqs.\ (\ref{tildetopen}) and (\ref{tildetclosed}). See Refs.\ 
\citep{Ratra1985, Ratra1989, Ratra1992, Ratra1994, RatraPeebles1995, 
Ratra2017} for discussions of such initial conditions in a variety of flat 
and non-flat inflation models. These initial conditions result in de Sitter
invariant inflaton two-point correlation functions in the very-slow-roll 
flat and closed de Sitter inflation models. The non-flat open and closed 
not-necessarily very-slow-roll inflation models which we apply them to here 
have less symmetry than the flat and closed de Sitter inflation models and so
it is possible that our assumption here of the absence of an additional
subdominant at large $A$ correction, that cannot be determined from Eq.\
(\ref{initialcondition}), and that might be important at small $A$, might
not be justified in the non-flat not-necessarily very-slow-roll inflation models
(such corrections do not contribute in the flat and closed de Sitter
inflation models).\footnote{I thank A.\ Guth for emphasizing this
point.} 
   
We have not been able to analytically compute the large $A$ asymptotic limit of 
Eq.\ (\ref{newscalarfieldperturbationsolution}).\footnote{Except for the 
$\hat{c}_-$ term in the closed case where the result agrees with the 
expression given in Eq.\ (\ref{closedc-}) below.} However, if we set 
$\hat{c}_+ = 0$ in both non-flat models, and choose for the closed model
\begin{eqnarray}
\label{closedc-}
\hat{c}_- = \left({16 \pi \over m_p{}^2}\right)^{1/2} {2^{-2A/p} (-1)^{A/p} \over \sqrt{2A}}, 
\end{eqnarray}
and for the open model
\begin{eqnarray}
\label{openc-}
\hat{c}_- = \left({16 \pi \over m_p{}^2}\right)^{1/2} {i 2^{-i2A/p}\over \sqrt{2A}},
\end{eqnarray}
it may numerically be shown in both non-flat models that the initial 
condition of Eq.\ (\ref{initialcondition}) is satisfied, \citep{Namjoo2020}.
Additionally, we show below that on smaller scales when spatial curvature is 
unimportant, the primordial power spectra in these tilted non-flat inflation 
models (which depend on these expressions for $\hat{c}_\pm$), are 
identical to the primordial power spectrum in the tilted flat inflation model 
of Refs.\ \citep{LucchinMatarrese1985, Ratra1992, Ratra1989}, where in 
Ref.\ \citep{Ratra1992} the corresponding asymptotic limits were computed.

\subsection{$P_{\mathcal{R}}$ at late time during inflation in the open and closed models}
  
Here we record expressions for the power spectra at late time during inflation 
in the open and closed tilted inflation models. The power spectrum definition 
and conventions are given in Appendices B and C. 

Using Eqs.\ (\ref{zetaPhi}) and (\ref{closedc-}), and setting $\hat{c}_+= 0$, 
we find at late time during inflation in the closed model  
\begin{eqnarray}
\label{PzetaClosed}
\sqrt{|P_{\mathcal{R}}|}  & = & \left({16 \pi \over m_p{}^2}\right)^{1/2} 
              \!\!\! Q^{1/p} { (2+q)p \over \sqrt{\pi q}} 
             \left|-1 + {W \over p}\right| \\
  \times  & {} & \!\!\!\!\!\! { 2^{-(6 -4q + 2A -W)/p} \over \sqrt{A} (A-1) (A+3)} 
            \left|{\Gamma\left(1 + W/p\right) \Gamma\left((2+q)/(2p)\right) \over \Gamma\left((2+W)/p\right) }\right|, \nonumber
\end{eqnarray}
where
\begin{eqnarray}
\label{Wcloseddef}
W = \sqrt{-8 - 4q + q^2 + 4 A (A+2)},
\end{eqnarray}
while  Eqs.\ (\ref{zetaPhi}) and (\ref{openc-}), and setting $\hat{c}_+= 0$, 
give at late time during inflation in the open model
\begin{eqnarray}
\label{PzetaOpen}
 \sqrt{|P_{\mathcal{R}}|} & = & \left({16 \pi \over m_p{}^2}\right)^{1/2} 
              \!\!\! Q^{1/p} { (2+q)p \over \sqrt{\pi q}}  
             \left|-1 + {W \over p}\right| \\
 \times & {} & \!\!\!\!\!\! { 2^{-(6 -4q)/p} \over \sqrt{A} (A^2 + 4)} 
            \left|{\Gamma\left(1 + W/p\right) \Gamma\left((2+q)/(2p)\right) \over \Gamma\left((2+W)/p\right) }\right|, \nonumber 
\end{eqnarray}
where
\begin{eqnarray}
\label{Wopendef}
W = \sqrt{-12 - 4q + q^2 - 4 A^2},
\end{eqnarray}
and in both cases the scalar spectral power-law index $n =(2 - 3q)/(2 -q)$. 
In the closed case, for fixed parameter values, the power spectrum of Eq.\ 
(\ref{PzetaClosed}) agrees very well with power spectra computed numerically 
in the model described by the potential energy density of Eq.\ 
(\ref{closedped}), \citep{Namjoo2020}. For representative plots of these 
and other non-flat inflation model power spectra, see Fig.\ 1 of Ref.\ 
\citep{deCruzetal2022}. 

In the large $A$ limit where $W = 2A$ ($2iA$) in the tilted closed (open) 
case, the $P_{\mathcal{R}}$ expressions in Eqs.\ (\ref{PzetaClosed}) and 
(\ref{PzetaOpen}) reduce identically to the flat-space tilted inflation
model expression of Eq.\ (\ref{PzetaEI}). This shows that on small scales at
late times during tilted non-flat inflation, when spatial curvature is not 
important, the tilted closed and open model primordial power spectra are 
identical to the tilted flat model primordial power spectrum. This is a 
useful consistency test of our analyses, including the initial conditions we have used here. This is because there is somewhat less uncertainty about initial conditions in the flat inflation model compared to the closed inflation model, as well as possibly even the open inflation model. Given that potentially only a tiny $\sim 1$\% deviation from flatness 
is what is of interest, it is reassuring that this initial conditions 
consistency test is passed.

\section{Conclusion}
\label{conclusion}

We have extended from the very-slow-roll, untilted, linear inflaton potential 
energy density open and closed inflation models of Refs.\ 
\citep{RatraPeebles1995, Ratra2017} to not-necessarily very-slowly-rolling, 
tilted, non-linear inflaton potential energy density open and closed 
inflation models. We have determined power spectra for quantum-mechanically
produced spatial inhomogeneities in these models. These power spectra  
can be used to characterize spatial inhomogeneities in closed and open 
inflation models, and have been used in analyses of CMB anisotropy and other
data, \citep{deCruzetal2022}. They differ from those that have previously
been used for this purpose, \citep{Planck2016, Planck2020}. The power spectra
of Refs.\ \citep{Planck2016, Planck2020} can also be generated by quantum 
fluctuations during (so far, only in closed) inflation, assuming different 
initial conditions, \citep{Guthetal2022}.     

Recent hints of observational tension with a few predictions of the 
spatially-flat $\Lambda$CDM model provides motivation for studying 
non-flat cosmological models as well as other alternatives. Moreover,
even if space is flat, to properly establish this from CMB anisotropy 
data requires use of consistent non-flat cosmological models --- such as 
those constructed here --- and the primordial power spectra in these models,
and in Ref.\ \citep{Guthetal2022} for the closed case. It might be significant
that in the non-flat inflation model there appears to be additional freedom
in the form of the inflation generated power spectrum, compared to the
spatially-flat case. 

\acknowledgements

I acknowledge very valuable discussions with A.\ Guth, M.\ H.\ Namjoo,
C.-G.\ Park, and J.\ de Cruz. This work was supported in part by DOE
grant DE-SC0011840.

\appendix

\section{${\mathcal{R}}_\Phi$ in the spatially-flat tilted inflation model}

In the spatially-flat tilted inflation model of 
Refs.\ \citep{LucchinMatarrese1985, Ratra1992, Ratra1989}, the scalar field 
potential energy density during inflation is
\begin{eqnarray}
\!\!\!\! V(\Phi) = \left({ 6-q\over 3} \right) {16 \pi \over m_p{}^2} 
               \rho^{(0)}_\Phi 
          {\rm exp} \left[ -\sqrt{q\over 2} (\Phi - \Phi^{(0)}) \right],
\end{eqnarray}
where $\Phi^{(0)}$ and $\rho^{(0)}_\Phi$ are the scalar field and the scalar 
field energy density during inflation at scale factor $a_0$. The scalar field 
energy density during inflation is
\begin{eqnarray}
    \rho_\Phi = \rho^{(0)}_\Phi \left({a_0 \over a}\right)^q.
\end{eqnarray}

During inflation, the gauge-invariant
\begin{eqnarray}
\label{zetaEI}
{\mathcal{R}}_\Phi & = & -{1 \over 6} \left({q \over 2}\right)^{(5-q)/q} {aH^{(10-q)/(2q)} \over k} \\
       & \times & \Big[ c_+ H^{(1)}_{\nu + 1} \left({2 k\over paH}\right) 
                   + c_- H^{(2)}_{\nu + 1} \left({2 k\over paH}\right)\Big], \nonumber 
\end{eqnarray}
where $H$ is the Hubble parameter, $\nu = (2+q)/(2p)$, $H^{(i)}_{\nu + 1}$ are 
Hankel functions, and from the initial conditions the constant of integration 
$c_- = 0$ and 
\begin{eqnarray}
& & c_+ =  \\
& &   \left({16 \pi \over m_p{}^2}\right)^{1/2} {k \over 2} 
      \left( { q \pi \over p} \right)^{1/2} \left(a_0 M^{2/q}\right)^{-5/2}
      e^{i(\nu - 1/2)\pi/2}, \nonumber
\end{eqnarray}
where
\begin{eqnarray}
M = {q \over 2} \left({8 \pi \over3 m_p{}^2} \rho_\Phi^{(0)}\right)^{1/2}.
\end{eqnarray}
 
\section{Relation between $P$ and ${\mathcal P}$ in flat, open, and closed models}

In this Appendix we define two power spectra we use and relate them to the 
two-point function in spatial momentum space. In this Appendix we do not 
explicitly indicate the time dependence of the fields.

In the flat model, defining the Fourier expansion of a position space field 
\begin{eqnarray}
\zeta (\vec {x}) = \int {d^3k\over (2 \pi)^3} e^{i \vec {k}\cdot \vec {x}} 
       \zeta (\vec {k}),
\end{eqnarray}
and 
\begin{eqnarray}
\langle \zeta (\vec {k}) \zeta^* (\vec {k'}) \rangle = P_\zeta(k) (2 \pi)^3 
     \delta^{(3)}(\vec {k} - \vec {k'}),
\end{eqnarray}
where $P_\zeta(k)$ is the power spectrum and $k = |\vec k|$, we have 
\begin{eqnarray}
& & \langle \zeta (\vec {x}) \zeta^* (\vec {x'}) \rangle = 
            \int {d^3k\over (2 \pi)^3} \int {d^3k'\over (2 \pi)^3} 
            e^{i ( \vec {k}\cdot \vec {x} - \vec {k'}\cdot \vec {x'})}
            \langle \zeta (\vec {k}) \zeta^* (\vec {k'}) \rangle \nonumber \\ 
& & \ \   = \int {d^3k\over (2 \pi)^3}  
            e^{i \vec {k}\cdot (\vec {x} - \vec {x'})} P_\zeta(k).
\end{eqnarray}
Setting $\vec{x} = \vec{x'}$, we have 
\begin{eqnarray}
\langle |\zeta (\vec {x})|^2 \rangle  = \int {d^3k\over (2 \pi)^3} P_\zeta(k)
= \int_0^\infty {d k \over k} {k^3 P_\zeta(k) \over 2 \pi^2},
\end{eqnarray}
and so define the power spectrum
\begin{eqnarray}
\label{scriptPflat}
\mathcal{P}_\zeta(k) = {k^3 P_\zeta(k) \over 2 \pi^2}
\end{eqnarray}
which gives the power in a logarithmic wavenumber interval.

In the open model, defining
\begin{eqnarray}
\zeta (\vec {\Omega}) = \int_0^\infty dA \sum_{BC} Z_{ABC} (\vec {\Omega}) 
    \zeta (A),
\end{eqnarray}
where $B$ and $C$ are `magnetic' integral indices, $Z_{ABC}$ is defined 
in Eq.\ (2.9) of Ref.\ \citep{RatraPeebles1995}, and 
\begin{eqnarray}
\langle \zeta (A) \zeta^* (A') \rangle = P_\zeta(A) \delta(A - A') \delta_{B,B'} 
\delta_{C,C'},
\end{eqnarray}
we have 
\begin{eqnarray}
& & \langle \zeta (\vec {\Omega}) \zeta^* (\vec {\Omega'}) \rangle \nonumber \\
& & \ \   = \int dA \int dA'  \sum_{BC}  \sum_{B'C'}
            Z_{ABC} (\vec {\Omega}) Z^*_{A'B'C'} (\vec {\Omega'})
            \langle \zeta (A) \zeta^* (A') \rangle \nonumber \\ 
& & \ \   = \int dA  \sum_{BC} Z_{ABC} (\vec {\Omega}) Z^*_{ABC} (\vec {\Omega'})
            P_\zeta(A) \nonumber \\
& & \ \   = \int_0^\infty {dA \over A} {A^3 P_\zeta(A) \over (2 \pi)^{3/2}}
            {P^{-1/2}_{iA -1/2} ({\rm cosh} (\gamma_3)) \over 
            \sqrt{{\rm sinh} (\gamma_3)}},
\end{eqnarray}
where we have used Eq.\ (2.11) of Ref.\ \citep{RatraPeebles1995}, $P^\mu_\nu$ 
is the associated Legendre function of the first kind, and 
${\rm cosh} (\gamma_3)$ is defined in Eq.\ (\ref{opengamma3}).
Setting $\vec{\Omega} = \vec{\Omega'}$ and using Eq.\ (A.3) of Ref.\ 
\citep{Ratra1994}, we have 
\begin{eqnarray}
\langle |\zeta (\vec {\Omega})|^2 \rangle   
= \int_0^\infty {d A \over A} {A^3 P_\zeta(A) \over 2 \pi^2},
\end{eqnarray}
and so define
\begin{eqnarray}
\label{scriptPopen}
\mathcal{P}_\zeta(A) = {A^3 P_\zeta(A) \over 2 \pi^2}.
\end{eqnarray}

In the closed model, defining
\begin{eqnarray}
\zeta (\vec {\Omega}) = \sum_{A = 2}^\infty \sum_{BC} Y_{ABC} (\vec {\Omega}) 
    \zeta (A),
\end{eqnarray}
where $B$ and $C$ are `magnetic' integral indices, $Y_{ABC}$ is defined in 
Eq.\ (9) of Ref.\ \citep{Ratra2017}, and 
\begin{eqnarray}
\langle \zeta (A) \zeta^* (A') \rangle = P_\zeta(A) \delta_{A,A'} \delta_{B,B'} 
\delta_{C,C'},
\end{eqnarray}
we have 
\begin{eqnarray}
& & \langle \zeta (\vec {\Omega}) \zeta^* (\vec {\Omega'}) \rangle \nonumber \\
& & \ \   = \sum_{A = 2}^\infty \sum_{A' = 2}^\infty  \sum_{BC}  \sum_{B'C'}
            Y_{ABC} (\vec {\Omega}) Y^*_{A'B'C'} (\vec {\Omega'})
            \langle \zeta (A) \zeta^* (A') \rangle \nonumber \\ 
& & \ \   = \sum_{A = 2}^\infty \sum_{BC} Y_{ABC} (\vec {\Omega}) 
            Y^*_{ABC} (\vec {\Omega'}) P_\zeta(A) \nonumber \\
& & \ \   = \sum_{A = 2}^\infty {1 \over A+1} {(A+1)^3 P_\zeta(A) \over (2 \pi)^{3/2}}
            {P^{-1/2}_{A -1/2} ({\rm cos} (\gamma_3)) \over 
            \sqrt{{\rm sin} (\gamma_3)}},
\end{eqnarray}
where we have used Eq.\ (11) of Ref.\ \citep{Ratra2017}, and 
${\rm cos} (\gamma_3)$ is defined in Eq.\ (\ref{closedgamma3}).
Setting $\vec{\Omega} = \vec{\Omega'}$ and using Eq.\ (3.9.2 (8)) on page 
163 of Ref.\ \citep{EHTF}, 
we have 
\begin{eqnarray}
\label{scriptPclosed}
\langle |\zeta (\vec {\Omega})|^2 \rangle 
= \sum_{A = 2}^\infty {1 \over A+1} {(A + 1)^3 P_\zeta(A) \over 2 \pi^2},
\end{eqnarray}
(the factor of $1/(A+1) = 1/\nu$ is chosen for consistency with the definition
of Ref.\ \citep{LesgourguesTram2014}, see below) and so define
\begin{eqnarray}
\mathcal{P}_\zeta(A) = {(A + 1)^3 P_\zeta(A) \over 2 \pi^2}.
\end{eqnarray}

\section{$P_{\mathcal{R}}$ and  ${\mathcal P_{\mathcal{R}}}$ at late time in the flat exponential potential and the open and closed linear potential inflation models, and comparison to CAMB and CLASS input power spectra}

In the flat model of Refs.\ \citep{LucchinMatarrese1985, Ratra1992, Ratra1989}, 
at late time during inflation, using the results of Appendix A and Eq. (9.1.9) 
of Ref.\ \citep{AS}, we have 
\begin{eqnarray}
\label{PzetaEI}
P_{\mathcal{R}} = C k^{n-4},
\end {eqnarray}
where $n = (2-3q)/(2-q)$ and the proportionality constant
\begin{eqnarray}
C = { 4 \over m_p{}^2 q} \Gamma^2\!\!\left({6 - q \over 2p}\right) p^{4/p} 
    \left({8 \pi \over3 m_p{}^2} \rho_\Phi^{(0)} a_0{}^q\right)^{2/p},
\end{eqnarray}
so, from Eq.\ (\ref{scriptPflat}),
\begin{eqnarray}
\label{Pzetaflat}
\mathcal{P_{\mathcal{R}}} = {C \over 2 \pi^2}k^{n-1}.
\end {eqnarray}
In the limit of small $q = 2 \epsilon^2$, Eq.\ (\ref{PzetaEI}) reduces to
\begin{eqnarray}
\label{PzetaEISR}
P_{\mathcal{R}} = {2 \pi h^2 \over m_p{}^2 \epsilon^2} k^{-3},
\end {eqnarray}
where $h^2 = 4 \pi (6-q) \rho_\Phi^{(0)}/(9 m_p{}^2)$. 

In the open linear scalar field potential energy density model of Ref.\ 
\citep{RatraPeebles1995}, with  $V(\Phi) = 12 h^2 [1 - \epsilon \Phi ]$ where 
$h$ is a constant, at late times during inflation
\begin{eqnarray}
\label{PzetaOI}
P_{\mathcal{R}} = {2 \pi h^2 \over m_p{}^2 \epsilon^2} {1 \over A (A^2 + 1)},
\end {eqnarray}
so
\begin{eqnarray}
\label{Pzetaopen}
\mathcal{P_{\mathcal{R}}} =  {h^2 \over \pi m_p{}^2 \epsilon^2}
{A^2 \over A^2 + 1}.
\end {eqnarray}

In the closed linear scalar field potential energy density model of Ref.\ 
\citep{Ratra2017}, at late times during inflation
\begin{eqnarray}
\label{PzetaCI}
P_{\mathcal{R}} = {2 \pi h^2 \over m_p{}^2 \epsilon^2} {1 \over A (A + 1) (A + 2)},
\end {eqnarray}
so
\begin{eqnarray}
\label{Pzetaclosed}
\mathcal{P_{\mathcal{R}}} =  {h^2 \over \pi m_p{}^2 \epsilon^2} {(A + 1)^2 \over A(A + 2)}.
\end {eqnarray}

In the limit $A \gg 1$, Eqs.\ (\ref{PzetaOI}) and (\ref{PzetaCI}) become
\begin{eqnarray}
P_{\mathcal{R}} = {2 \pi h^2 \over m_p{}^2 \epsilon^2} A^{-3},
\end {eqnarray}
which is identical to the expression in Eq.\ (\ref{PzetaEISR}). As expected, 
on small scales at late times during non-flat inflation when spatial curvature 
is unimportant, the very-slow-roll closed and open inflation model primordial 
power spectra are identical to the very-slow-roll flat inflation model 
primordial power spectrum.   

The CAMB and CLASS input power spectra are defined in Ref.\ 
\citep{LesgourguesTram2014}. The definition in their Eq.\ (3.23) is a 
little unusual given the $(2 \pi)^3$ in their Eq.\ (3.24), but the 
normalization of the power spectrum is an adjustable parameter to be 
determined by fitting to data.  

Comparing Ref.\ \citep{LesgourguesTram2014} Eqs.\ (3.25) and (3.26) in the 
flat case, we see that their Eq.\ (3.26) is identical to Eq.\ 
(\ref{Pzetaflat}) above.

Reference \citep{TramLesgourgues2013} defines the negative of the eigenvalue of 
the spatial Laplacian to be $k^2/|K|$ in their Eq.\ (1.11). Here 
$K = - H_0^2 \Omega_{k0}$ is negative (positive) for open (closed) 
spatial hypersurfaces and $\Omega_{k0}$ is the current value of the spatial 
curvature density parameter. In the second line below Eq.\ (3.4) of Ref.\ 
\citep{LesgourguesTram2014} (where they define $q$ in terms of $k$) they 
define another wavenumber $\nu = q/\sqrt{|K|}$. It can be seen that in 
the open model their $\nu$ is identical to the $A$ we use here while in 
the closed model their $\nu$ is identical to the $A + 1$ we use here.

Reference \citep{LesgourguesTram2014} uses an unusual but now standard 
convention. 
They define  $\mathcal{P_{\mathcal{R}}} = A_s k^{n-1}$ to be the flat space 
expression, see
their Eq.\ (3.26). To make this clear, in what follows, we put a superscript 
${\rm FS}$ on this and write $\mathcal{P^{\rm FS}_{\mathcal{R}}} = A_s k^{n-1}$. 
What we call $\mathcal{P_{\mathcal{R}}}$ they refer to as 
\begin{eqnarray}
\widetilde{\mathcal{P}}_{\mathcal{R}} (\nu) = {\nu^2 \over \nu^2 - \hat{K}} 
\mathcal{P^{\rm FS}_{\mathcal{R}}}.
\end{eqnarray}
Here $\hat{K} = K/|K|$ which is $-1(+1)$ for open (closed) hypersurfaces. The 
above expression can be derived from their Eq.\ (3.29) by changing variables 
from $q$ to $\nu$. That our $\mathcal{P_{\mathcal{R}}}$ is their 
$\widetilde{\mathcal{P}}_{\mathcal{R}}$ follows from their Eq.\ (3.28) by 
rewriting $d\nu \, \nu^2$ as  $(d\nu /\nu) \, \nu^3$ and combining the $\nu^3$
with the last factor in the integrand and then comparing to the flat-space 
expression in their Eq.\ (3.25).  

Setting $n=1$ in the  Ref.\ \citep{LesgourguesTram2014} non-flat expressions, 
as we want to compare to the linear scalar field potential energy density 
inflation expressions above, we find in the open case the Ref.\ 
\citep{LesgourguesTram2014} formula is
\begin{eqnarray}
\mathcal{P_{\mathcal{R}}} \propto {A^2 \over A^2 + 1},
\end {eqnarray}
that is in agreement with Eq.\ (\ref{Pzetaopen}) above, and in the closed 
case the Ref.\ \citep{LesgourguesTram2014} formula is 
\begin{eqnarray}
\mathcal{P_{\mathcal{R}}} \propto {(A + 1)^2 \over A(A + 2)},
\end {eqnarray}
which agrees with Eq.\ (\ref{Pzetaclosed}) above.

\end{document}